\documentclass{PoS}
\usepackage{amsmath}
\usepackage{dsfont}
\usepackage{slashed}
\usepackage{booktabs}
\usepackage{graphicx}
\usepackage{float}
\floatstyle{plain} 
\restylefloat{figure} 
\title{Omega baryon electromagnetic form factors from lattice QCD}

\ShortTitle{Omega baryon electromagnetic form factors from lattice QCD}

\author{C. Alexandrou\\
          Department of Physics, University of Cyprus, P.O. Box 20537, 1678 Nicosia, Cyprus and\\
      Computation-based Science and Technology Research Center, The Cyprus Institute, P.O. Box 27456, 1645 Nicosia, Cyprus\\
         E-mail:           \email{alexand@ucy.ac.cy}}

\author{T. Korzec\footnote{Current address:Institut f\"ur Physik
   Humboldt Universit\"at zu Berlin, Newtonstrasse 15, 12489 Berlin, Germany.} \\
        Department of Physics, University of Cyprus, P.O. Box 20537, 1678 Nicosia, Cyprus\\
        E-mail: \email{korzec@ucy.ac.cy}}

\author{G. Koutsou\\
        Department of Physics, University of Wuppertal/Forschungszentrum J\"{u}lich D-52425, J\"{u}lich, Germany\\
        E-mail: \email{i.koutsou@fz-juelich.de}}

\author{\speaker{Y. Proestos}\\%
     Computation-based Science and Technology Research Center, The Cyprus Institute, P.O. Box 27456, 1645 Nicosia, Cyprus\\
       E-mail: \email{y.proestos@cyi.ac.cy}}

\abstract{We present results on the Omega baryon ($\Omega^{-}$) electromagnetic form factors
using  $N_f=2+1$ dynamical domain-wall fermion configurations 
corresponding to a pion mass of about 330 MeV.  
We construct  appropriate
sequential sources for the determination of the
 two dominant form factors $G_{E0}$ and $G_{M1}$ as well as a sequential
source that isolates the subdominant electric quadrupole form factor $G_{E2}$.
We calculate the $\Omega^-$ magnetic moment, $\mu_{\Omega^{-}}$, and 
 electric charge radius, $\langle r^{2}_{E0} \rangle$, 
and compare to experiment, for the case of $\mu_{\Omega^{-}}$,  and to other lattice calculations. 
}

\FullConference{The XXVII International Symposium on Lattice Field Theory\\
                July 26-31, 2009\\
                Peking University, Beijing, China}

\begin{document}

\section{Introduction}
The structure of hadrons, such as size, shape and charge distribution 
can be probed by the electromagnetic form factors. 
The $\Omega^{-}$ baryon, consisting of three valence strange
quarks, is significantly more stable than other members of the baryon decuplet,
 such as the $\Delta$, with a life-time on the order of $10^{-10}s$. 
This fact makes the calculation of its electromagnetic form factors  
particularly interesting since they are accessible to experimental 
measurements. 
Its   magnetic dipole moment is measured to very good
accuracy unlike those of the other decuplet baryons. A value of 
$\mu_{\Omega^{-}}=-2.02(5)$ 
is given in the PDG~\cite{PDG:2008}  in  units of nuclear magnetons ($\mu_N$).
 Within lattice QCD one can directly compute hadron form factors 
starting from the fundamental theory of the strong interactions. 
 Furthermore, higher order multipole moments, not detectable by current experimental setups, are accessible to lattice methods. 
Higher-order moments such as the electric quadrupole are essential in 
the determination of the deformation of a hadron state. 
 In this work, we calculate the electromagnetic form factors of 
the $\Omega^{-}$ baryon using, for the first time, dynamical 
domain-wall fermion configurations. For the calculation we use 
the fixed-sink approach which enables us to calculate the form
 factors for all values and directions of the momentum 
transfer $\vec q$ concurrently. The main advantages of this approach is that it allows an increased statistical precision, while at the same time it provides  the full $Q^2$ dependence. Moreover, 
by constructing optimized sources we  calculate
 the two dominant form factors accurately.
An appropriately defined source for the subdominant electric
 quadrupole form factor is constructed and tested~\cite{Alexandrou:2008bn}. 
The aim is to obtain an accurate determination
of the quadrupole moment, at a price, of course, of additional sequential inversions.

\section{Electromagnetic form factors of the $\Omega^{-}$ baryon}
The on-shell $\Omega^{-}$ matrix element of 
the electromagnetic current $j_{EM}^{\mu}$, can be decomposed in terms of four independent Lorentz covariant vertex functions, $a_1(q^2)$, $a_2(q^2)$, $c_1(q^2)$ and $c_2(q^2)$, which  depend only on the squared momentum transfer $q^2=-Q^2=(p_{i}-p_{f})^2$. In Minkowski spacetime these are given  by~\cite{Nozawa:1990gt}
 \begin{align}\label{matrelem}
 \langle \Omega(p_f,s_f) | j_{\rm EM}^\mu | \Omega(p_i,s_i)\rangle =  \sqrt{\frac{m^{2}_{\Omega}}{E_\Omega(\vec{p_f})  E_\Omega(\vec{p_i})}}
&
\,\bar{u}_\sigma(p_f,s_f){\cal O}^{\sigma \mu \tau} u_\tau(p_i,s_i),\\[1mm]
\mathcal{O}^{\sigma \mu \tau} =-g^{\sigma \tau}\biggl[a_1(q^2) \gamma^\mu +\frac{a_2(q^2)}{2m_\Omega} \left(p_f^\mu + p_i^\mu\right)\biggr]
& -\frac{q^\sigma q^\tau}{4m_\Omega^2}\biggl[c_1(q^2)\gamma^\mu + \frac{c_2(q^2)}{2m_\Omega}\left(p_f^\mu+p_i^\mu\right)\biggr].
\end{align}
The rest mass and the energy of the particle are denoted by $m_\Omega$ and $E_\Omega.$ The initial (final) four-momentum and spin-projection are given by $p_{i}\, (p_{f})$ and $s_{i}\, (s_{f})$. In addition, every vector component of the spin-$\frac{3}{2}$ Rarita-Schwinger vector-spinor $u_{\sigma}$ 
satisfies the Dirac equation,
$   \big(p_{\mu}\gamma^{\mu}-m_{\Omega}\big) u^\sigma(p,s) = 0$,   
along with the auxiliary conditions: $  \gamma_\sigma u^\sigma(p,s) = 0$ and $ p_\sigma u^\sigma(p,s)= 0$.
Furthermore, the covariant vertex functions are linearly related to the experimentally measured electric  $G_{E0}(q^2)$, $G_{E2}(q^2)$ and magnetic $G_{M1}(q^2)$ and $G_{M3}(q^2)$ multipole form factors~~\cite{Alexandrou:2008bn, Nozawa:1990gt}.

\section{Lattice techniques}
%
%
We employ gauge configurations generated by the RBC-UKQCD joint collaborations
using $N_f=2+1$ dynamical domain-wall fermions (DWF) and the Iwasaki gauge-action corresponding to a pion mass of about 330~MeV. The simulation
is carried out on a lattice
of size $24^3\times 64$ with a lattice spacing of 0.114(2)~fm.
For this  pion mass and  finite lattice volume the $\Omega^{-}$ is  stable.
The lattice spacing $a$, the light u and d  and  the strange quark mass were
fixed by an iterative procedure using the $\Omega^-$, the pion and the kaon masses~\cite{Allton:2008pn}. 
For the present calculations we use  200 well separated 
dynamical domain-wall fermion gauge configurations~\cite{Allton:2008pn}.
It is known that the chiral symmetry breaking falls exponentially
with the length of the fifth dimension $N_5$ of the DWF-action. The value $N_5=16$ used here is adequate to keep the residual mass sufficiently small.
\subsection{Interpolating fields}
In order to calculate the $\Omega^-$ matrix element
we need to evaluate the appropriate 
two- and three-point correlation functions.
An interpolating field operator with the quantum numbers of the $\Omega^{-}$ hyperon is given by
\begin{align}\label{interpolator1}
J_{\Omega}(x)&= \epsilon^{abc}\, \mathbf{s}^{a}_{\alpha}\,\big(\mathbf{s}^{\!\mbox{\tiny T}b}_{\beta}\, [C\gamma_{\sigma}]_{\beta\gamma}\,  \mathbf{s}^{c}_{\gamma} \big),
\end{align}
where $C=\gamma_4 \gamma_2$ is the charge-conjugation matrix and $\sigma$ is the vector index of the spin-3/2 spinor.

\vspace*{0.15cm}

\noindent
\begin{minipage}{0.5\linewidth}
In order to ensure  ground state dominance for the shortest
 time evolution  we perform a gauge invariant Gaussian
 smearing, as described in Refs.~\cite{Alexandrou:1992ti,APEsmearing}, on the quark fields $\mathbf{s}$ that enter in the interpolating field:
{\small
$$        {\mathbf q }_\beta(t,\vec x) = \sum_{ \vec y} [\mathds{1} + \alpha H(\vec x,\vec y; U)]^n \ q_\beta(t,\vec y),$$
$$	H(\vec x, \vec y; U)  = \sum_{\mu=1}^3 \left(U_\mu(\vec x,t)\delta_{\vec x, \vec y - \hat \mu} + U^{\dagger}_\mu(\vec x-\hat \mu, t) \delta_{\vec x,\vec y+\hat\mu} \right),$$}
\noindent
where $\mathbf{q}$ (q) is the smeared (local) fermion field. The links $U_{\mu}(\vec x,\, t)$ entering the hopping matrix $H$ 
are APE-smeared gauge fields.
\end{minipage}\hfill
\begin{minipage}{0.46\linewidth}
\includegraphics[width=\linewidth]{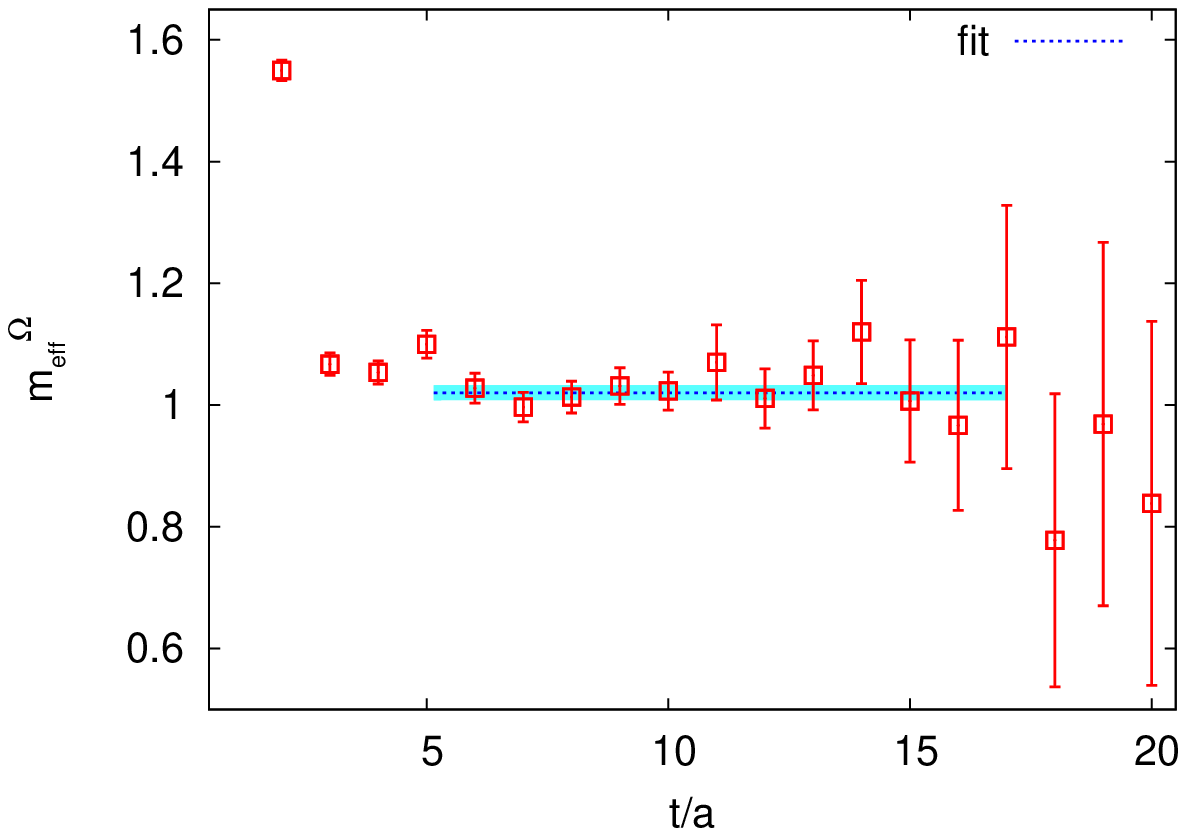}
Figure~1: {The $\Omega^-$ effective mass and the fit to the plateau plotted against time separation.}
\label{meff}
\end{minipage}\vspace*{0.45cm}

For the lattice spacing considered here we have used the Gaussian smearing parameters $\alpha=5.026$ and $n=40$, which have  been optimized for the nucleon state. In Fig.~1 we show
the $\Omega^-$ effective mass, calculated from the two-point function ratio defined by $m_{eff}^{\Omega}(t)=-\log[G(t+1,{\vec{0}})/G(t,\vec{0})]$. It displays a nice plateau yielding $m_{\Omega}= 1.76(2) \ \mathrm{GeV}$. This value
is 5\% higher than the  experimental one. This may reflect a slightly larger value for the strange quark mass as compared to the physical one.  
\stepcounter{figure}
\subsection{Two- and three-point Correlation functions}
The electromagnetic form factors are extracted by constructing appropriate combinations of two- and three-point correlation functions. The corresponding lattice correlation functions are given by
\begin{align}\label{twopointF}
G(t_{f},\vec q) &= \sum_{\vec x_f}\sum_{j=1}^3 e^{-i\vec x_f \cdot \vec q}\, 
       \Gamma^4_{\alpha\beta}   \langle J_{j\beta}(x_f)  \overline J_{j\alpha}(0) \rangle, \\[0.5mm]
G^{\ \mu}_{\sigma\ \tau}(\Gamma^\nu,t,\vec q) &= \sum_{\vec x_f \vec x} e^{i\vec x \cdot \vec q} 
       \Gamma^\nu_{\alpha\beta} \langle J_{\sigma\beta}(x_f) j_{EM}^\mu(x) \overline J_{\tau\alpha}(0) \rangle, 
\label{threepointF}
\end{align}
where  $\Gamma^4= \frac{1}{4}\big(\mathds{1}+\gamma^4\big)$ and $\Gamma^k = i\Gamma^{4}\gamma^5\gamma^k$, with the Dirac $\gamma$-matrices taken in Euclidean spacetime. 
We calculate the above three-point correlation function
  in a frame where the final $\Omega^{-}$ 
state is at rest, $\vec p_f=\vec 0$, while the operator
$j^\mu_{EM}$ is inserted at time $t$ carrying a momentum $\vec q=-\vec p_i$.

The leading time dependence and unknown overlaps of the $\Omega^-$
 state with the initial state
$\bar{J}_{\Omega}|0\rangle$ 
 in the three-point correlation function can be canceled out by forming appropriate ratios of the three-point function and two-point functions. A particularly
suitable ratio is defined by
\begin{align}\label{ratio1}
        R_{\sigma\ \tau}^{\ \mu}(\Gamma,\vec q,t) &= \frac{G_{\sigma\ \tau}^{\ \mu}(\Gamma^{\nu},\vec q,t)}{G_{k k}(\Gamma^4,\vec 0, t_f)}\ 
				         \sqrt{\frac{G_{kk}(\Gamma^4,\vec p_i, t_f-t)G_{kk}(\Gamma^4,\vec 0  ,t)G_{kk}(\Gamma^4,\vec 0,t_f)}
					            {G_{kk}(\Gamma^4,\vec 0, t_f-t)G_{kk}(\Gamma^4,\vec p_i,t)G_{kk}(\Gamma^4,\vec p_i,t_f)}}\, ,
\end{align}
where a summation over the repeated indices $k \,\,(k=1,2,3)$ is assumed. For large Euclidean time separations this ratio becomes time-independent (plateau region)
\begin{align}\label{ratio2}
   R_{\sigma\ \tau}^{\ \mu}(\Gamma,\vec q,t) 
\stackrel{t_f-t \gg 1,t \gg 1}{\longrightarrow} \Pi_{\sigma\ \tau}^{\ \mu}(\Gamma,\vec q) &= \mathcal{C}\
  \mathrm{Tr}\left[\Gamma\, \Lambda_{\sigma\sigma }(p_f) \mathcal{O}^{{\sigma }\mu{\tau }} \Lambda_{\tau  \tau}(p_i) \right] ,
\end{align}
\vspace*{-5mm}
\begin{align}\label{prefactorratio}
\mathcal{C}&= \sqrt{\frac{3}{2}}\left[\frac{2 E_{\Omega}(\vec q)}{m_\Omega} 
                          +\frac{2 E^2_{\Omega}(\vec q)}{m^2_\Omega} 
                          +\frac{  E^3_{\Omega}(\vec q)}{m^3_\Omega} 
                          +\frac{  E^4_{\Omega}(\vec q)}{m^4_\Omega} \right]^{-\frac{1}{2}}. 
\end{align} 
It is understood that the trace acts on spinor-space, while the Euclideanized version of the Rarita-Schwinger spin sum is given by
\begin{align}\label{RSeuclidean}
   \Lambda_{\sigma\tau}(p)&\equiv \sum_{s} u_\sigma(p,s) \bar{u}_\tau(p,s) = - \frac{-i\slashed{p}+m_\Omega}{2m_\Omega}\left[
   \delta_{\sigma\tau}-\frac{\gamma_\sigma\gamma_{\tau}}{3}
                               +\frac{2p_\sigma p_{\tau}}{3m_\Omega^2} 
- i \frac{p_\sigma\gamma_{\tau}-p_{\tau}\gamma_\sigma}{3m_\Omega} \right] \, .
\end{align}

The electromagnetic form factors are extracted by fitting to the plateau region
of  a set of 
specially chosen  combinations of three-point functions 
given below
\begin{align}\label{optimalcombs1}
\sum_{k=1}^3                    \Pi_{k\ k}^{\ \mu}(\Gamma^4,\vec q) &= K_1\ G_{E0}(Q^2) + K_2\ G_{E2}(Q^2),\\[-1mm]
\sum_{j,k,l=1}^3 \epsilon_{jkl} \Pi_{j\ k}^{\ \mu}(\Gamma^4,\vec q) &= K_3\ G_{M1}(Q^2),  \label{optimalcombs2}\\[-1mm]
\sum_{j,k,l}^3   \epsilon_{jkl} \Pi_{j\ k}^{\ 4}(\Gamma^j,\vec q)   &= K_4\ G_{E2}(Q^2) \label{optimalcombs3} .
\end{align}
The continuum  kinematical coefficients $K_{i}\ \ (i=1,2,3,4),$
have been calculated in Ref.~\cite{Alexandrou:2009hs} and are functions 
of the $\Omega^{-}$ mass, the energy $E_{\Omega}$,
 the space-time index $\mu$ and the momentum $\vec q$. 
Furthermore, it is readily seen that the subdominant electric 
quadrupole form factor $G_{E2}$ is isolated by the combination 
provided by Eq.~(\ref{optimalcombs3}). In addition, these expressions are such that all possible directions 
of $\mu$ and momentum $\vec q$ contribute in a symmetric fashion at 
a given momentum transfer $Q^2.$ In other words, the optimal 
combinations of $\Pi_{\sigma\ \tau}^{\ \mu}(\Gamma,\vec q)$ employed 
are those that maximize the number of nonzero $\vec q$ contributions in a
 lattice rotationally invariant way~\cite{Alexandrou:2007dt}. 

We calculate only the connected contributions to the three-point function by
 performing sequential inversions through the sink. This means that we 
need to fix the quantum numbers of the initial and final states
and explains  why we consider optimal combinations.
Otherwise, one would have to prepare a new set of sequential inversions 
for every choice of vector and Dirac indices, which is prohibited in 
a lattice computation given the fact that we have to consider an overall of $\sigma \times \beta \times \tau \times \alpha =256$, combinations by looking the index structure of Eq.~(\ref{threepointF}). 
Having the matrix element for all the different directions of $\vec q$ 
and for all four directions $\mu$ of the current, 
 we form an over-constrained system of linear equations 
(in terms of form factors), which is solved by employing
 a \emph{singular value decomposition} analysis. 
This procedure yields
 the electromagnetic form factors $G_{E0}$, $G_{M1}$ and $G_{E2}$. 
The statistical errors are found by a jack-knife analysis, which takes care
 of the correlations of lattice measurements of the ratios.

\begin{figure}[htb]
\begin{center}
\hspace{-0.2cm}\includegraphics[width=0.7\linewidth,height=0.7\linewidth]{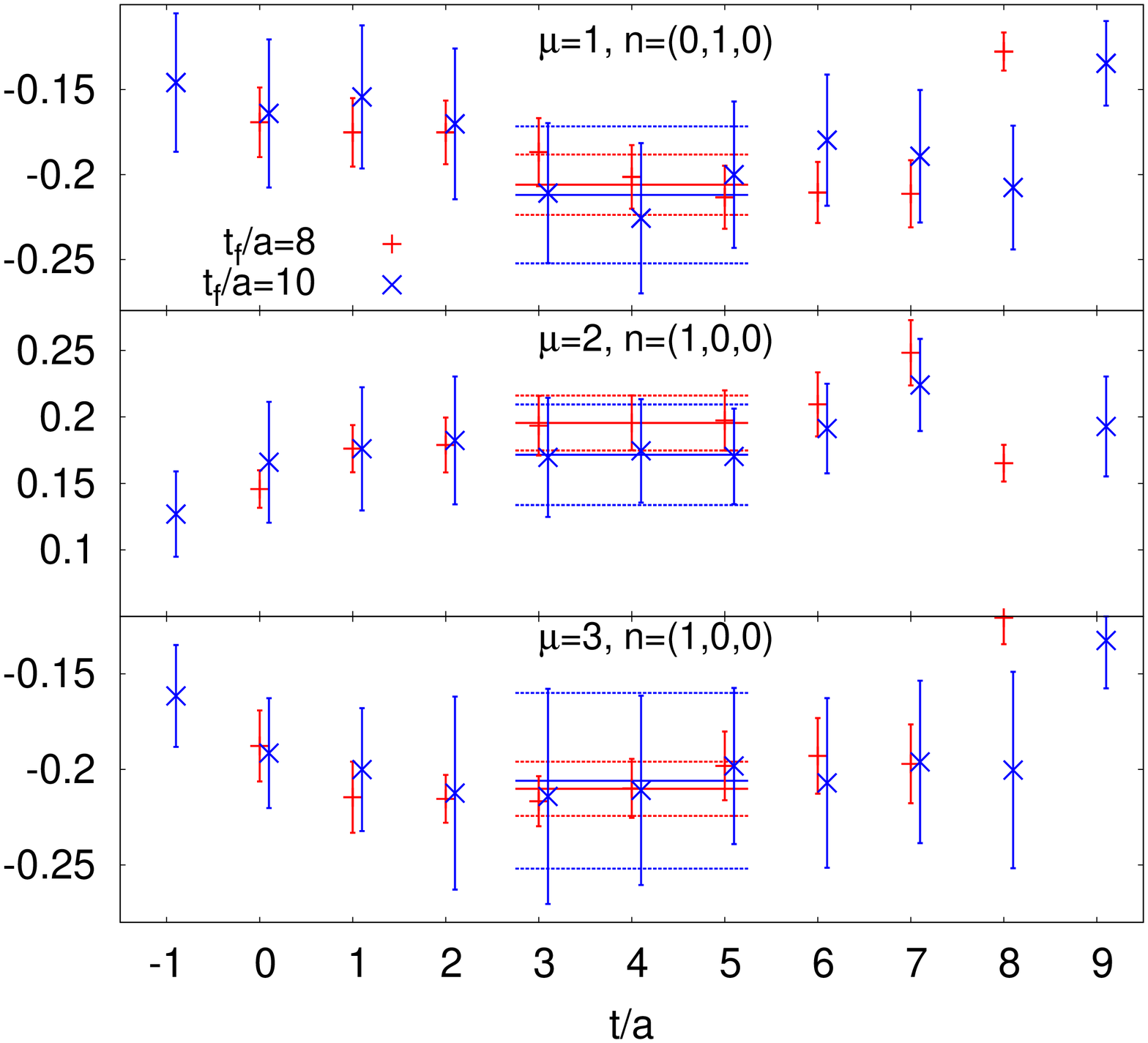}
\vspace*{-1cm}\caption{The ratio $R_{\sigma\ \tau}^{\ \mu}(\Gamma,\vec q,t)$
 extracted for temporal source-sink separations $t_f/a=8$ and $t_f/a=10$, 
using 50 gauge configurations.   The results for $t_f/a=10$ 
are shifted to the left by one unit. 
We show results for current directions $\mu=1,2,3$ 
and low-momenta $\vec q$: $(0,1,0)\frac{2\pi}{L}$ and $(1,0,0)\frac{2\pi}{L}$, respectively.
}
\label{figRRtype2}
\end{center}
\end{figure}

As already mentioned
 the three-point function of the connected part  is calculated by
 performing sequential inversions through the sink. This requires to fix the  
temporal source-sink separation. In order to determine
the smallest time separation that is still sufficiently large to damp
the excited state contributions we perform the calculation at two values of
the sink-source separation, namely $t_f/a=8$ and $t_f/a=10$. We  compare in
 Fig.~\ref{figRRtype2}
 the  results for the plateaus   $\Pi_{\sigma\ \tau}^{\ \mu}(\Gamma,\vec q)$,
 for a few selected directions of the current and for low momentum $\vec q$
 values for these two sink-source time separations. As can be seen, the
plateaus values at $t_f/a=10$ are consistent with the smaller time separation
having about half the statistical error. We therefore use $t_f/a=8$ or
$t_f=0.91$~fm  in what follows.

\section{Results}
We use the local electromagnetic current,  $j_{EM}^{\mu}=-\frac{1}{3}\bar{s}\gamma^{\mu}s$, which requires a renormalization factor $Z_{V}$ 
to be included. This renormalization constant is determined 
by the requirement that
$G_{E0}$ at zero momentum transfer is equal to the charge of $\Omega^{-}$ 
in units of electric charge, that is 
$G_{E0}(0)\equiv -1$.
Our calculation, at this quark mass, yields a value of 
$Z_{V}=0.727(1)$, which is 
reasonably close to  the value 
 obtained in Refs.~\cite{Aoki:2007xm,Allton:2008pn} 
in the chiral limit.
\subsection{Electric charge form factor}
\begin{figure}[htb]
\begin{center}
    \begin{tabular}{cc}
\hspace{-0.65cm}\includegraphics[width=0.52\linewidth,height=0.45\linewidth]{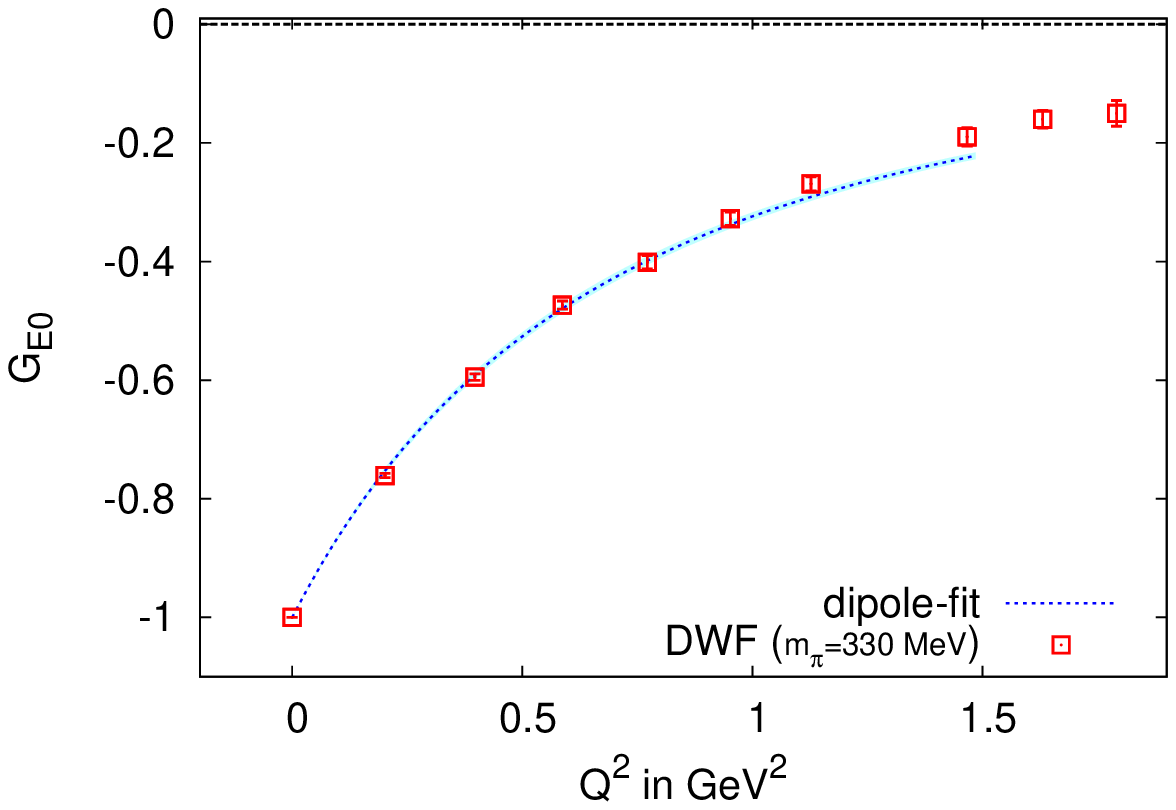} & \hspace{-0.15cm}\includegraphics[width=0.52\linewidth,height=0.45\linewidth]{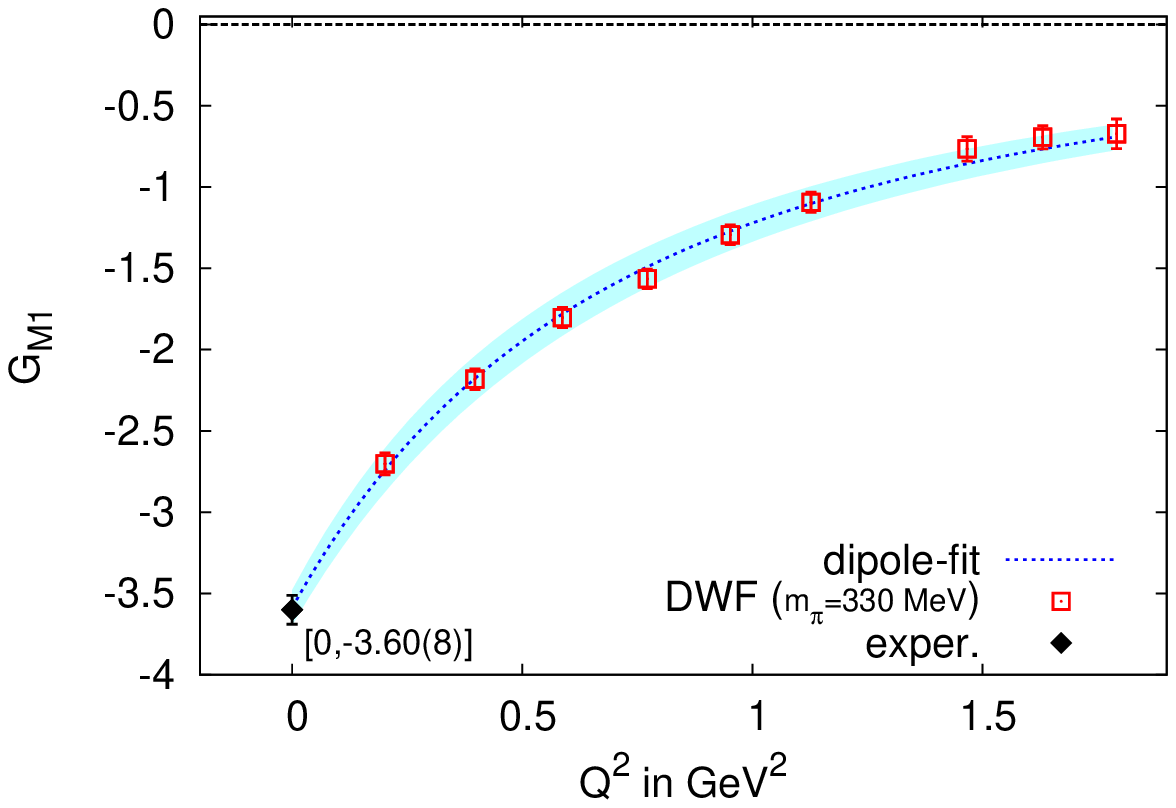} \\
   \qquad (a)& \qquad (b)\\[-1mm]
	\end{tabular}
\vspace*{-6mm}
   \caption{(a) The electric charge form factor $G_{E0}$ at  $m_{\pi}=330\ \mathrm{MeV}.$ The line describes the dipole fit given in Eq.~(\protect \ref{ge0dipoleform}). 
            (b) The magnetic dipole form factor, $G_{M1}$, along with the dipole fit as given in  Eq.~(\protect \ref{gm1dipoleform}) at $m_{\pi}=330\ \mathrm{MeV}$. The experimental datum for the magnetic dipole form factor, $G_{M1}(0)=-3.60(8)\ \mathrm{GeV}^2$~\cite{PDG:2008}, is also shown.
	    }\label{figge0gm1}
\end{center}
\end{figure}

Our results for the electric charge form factor, $G_{E0}$, are depicted in Fig.~\ref{figge0gm1}(a).
As can be seen, the momentum dependence of this form factor is  described well by a dipole form
\begin{align}\label{ge0dipoleform}
   G_{E0}(Q^2) &= -\frac{1}{\big(1+ \frac{Q^2}{\Lambda_{E0}^{2}}\big)^2} , 
\end{align}  
with $\Lambda_{E0}^{2}$ a fit parameter. 
In the non-relativistic limit the slope of the above dipole form evaluated 
at momentum transfer $Q^2=0$, is related to the electric charge 
root mean square (\emph{rms})  radius
\begin{align}\label{rmrradius}
   \left\langle r^{2}_{E0} \right\rangle = -6  \frac{d}{dQ^2} G_{E0}(Q^2)\bigg|_{Q^2=0} \, .
\end{align}
From our dipole fit to the  lattice data we determine $\Lambda_{E0}$ and
obtain
 a value of $\langle r^{2}_{E0}\rangle=-0.354(9)\ \mathrm{fm}^2$. 
This is slightly greater in magnitude than the one 
reported in Ref.~\cite{Boinepalli:2009sq}, which was obtained in
 a quenched lattice QCD calculation (see Table~\ref{ResultsTable}). 
Our value is expected to be higher since  in a dynamical 
 lattice calculation  meson-cloud effects are taken into account in addition to the fact that in Ref.~\cite{Boinepalli:2009sq} a heavy pion mass has been used.
\subsection{Magnetic dipole form factor}
Lattice results on the magnetic dipole $G_{M1}$ are shown in Fig.~\ref{figge0gm1}(b).
 As in the case of $G_{E0}$, a dipole fit 
describes very well the $Q^2$-dependence of $G_{M1}$. Fitting to the form
\begin{equation}
   G_{M1}(Q^2) = -\frac{G_{M1}(0)}{\big(1+ \frac{Q^2}{\Lambda_{M1}^{2}}\big)^2}
\label{gm1dipoleform}
\end{equation}
we can extract the anomalous magnetic moment of the $\Omega^-$.
By utilizing the lattice computed $\Omega$  mass, 
from Table~\ref{ResultsTable}, we calculate the  magnetic moment
in nuclear magnetons by using the  relation
\begin{align}\label{magmoment}
   \mu_{\Omega^{-}}&= G_{M1}(0) \frac{e}{2m_{\Omega}}=G_{M1}(0)\ \frac{m_N}{m_{\Omega}}\ \mu_N.
\end{align} 
Our value of $\mu_{\Omega^{-}}$ in nuclear magnetons $\mu_N$ is given in
 Table~\ref{ResultsTable} and it is in agreement
with the experimental value. It is also in agreement
with  two recent lattice calculations~\cite{Boinepalli:2009sq,Aubin:2008qp}.
The calculation in Ref.~\cite{Boinepalli:2009sq} is similar to ours
in the sense that the three-point correlation function is calculated but 
the evaluation is
carried out in the quenched theory and only at one value of $Q^2$.
 In Ref.~\cite{Aubin:2008qp} one employs a background field method to compute
energy shifts using $N_F=2+1$ Clover fermions at 
pion mass of 366~MeV on an anisotropic lattice.

\begin{table}[ht]
  \centering
    \begin{tabular}{l c c c c c c c}
		\toprule
		\toprule
 & Vol. & $N_{\mathrm{conf.}}$ & $m_\pi$ & $m_\Omega$ & $G_{M1}(0)$ &$\mu_{\Omega^{-}}$ & $\langle r^{2}_{E0}\rangle $  \\
 & &  & {\scriptsize [MeV] }& {\scriptsize [GeV] }& {\scriptsize [$\mathrm{GeV}^2$] }& {\scriptsize [$\mu_{N}$] } & {\scriptsize  [$\mathrm{fm}^{2}$] }\\
\midrule
this work & $24^{3}\times 64$ & 200 & 330   & 1.763(21) & -3.58(10) & -1.92(6) & -0.354(9)\\
ref.~\cite{Boinepalli:2009sq} & $20^{3}\times 40$ & 400 & 697   &  1.732(12) & -- & -1.697(65) & -0.307(15)   \\
ref.~\cite{Aubin:2008qp} & $24^{3}\times 128$ & 213 & 366   & 1.650 & -- & -1.93(8) & --   \\
\midrule
ref.~\cite{PDG:2008} {\scriptsize [PDG]} & -- & -- & -- & 1.672(45) & -3.60(8) & -2.02(5) &--   \\
%
\bottomrule\\
\end{tabular}
\vspace*{-6mm}
\caption{The $\Omega^{-}$ mass $m_{\Omega}$, the magnetic dipole form factor 
$G_{M1}$ at $Q^2=0$, the magnetic moment $\mu_{\Omega^{-}}$ 
and  the  electric {\it rms} charge radius $\langle r^{2}_{E0}\rangle$.
} \label{ResultsTable}
\end{table}
\section{Summary}
Using appropriately constructed sequential sources the dominant $\Omega^{-}$ electromagnetic form factors $G_{E0}$ and $G_{M1}$ are calculated  with good accuracy even with a small sample  of dynamical domain-wall fermion configurations. In the current calculation we neglected disconnected contributions.

The $\Omega^{-}$ magnetic moment is extracted by fitting the magnetic dipole form factor $G_{M1}$
 to a dipole form. We find a value that is in agreement 
with experiment~\cite{PDG:2008}.
The electric {\it rms} charge radius $\langle r^{2}_{E0}\rangle $ is 
also computed and found to be larger
than the value obtained in quenched QCD~\cite{Boinepalli:2009sq}.

We have also preliminary results on the subdominant electric quadrupole 
form factor $G_{E2}$ using the source of Eq.~(\ref{optimalcombs3}) but
with our current statistics the errors are still large
and no definite conclusion can be drawn.

We will check for cut-off effects
by performing the calculation of the
form factors  using dynamical domain-wall fermion
configurations at a finer lattice spacing. Although the light quark
mass dependence is expected to be small, this needs to be checked at another, 
preferably smaller, value of the pion mass.

\acknowledgments
This research was partly supported by the Cyprus Research Promotion Foundation (R.P.F) under contracts $\mathrm{\Pi}$ENEK/ENI$\mathrm{\Sigma}$X/0505-39 and EPYAN/0506/08.


\begin{thebibliography}{99}
 \bibitem{PDG:2008}
  C.~Amsler \emph{et al}. (Particle Data Group), PL B\textbf{667} (2008) 1.

\bibitem{Alexandrou:2008bn}
  C.~Alexandrou {\it et al.},
  Phys.\ Rev.\  D {\bf 79} (2009) 014507.



\bibitem{Nozawa:1990gt}
  S.~Nozawa and D.~B.~Leinweber,
  Phys.\ Rev.\  D {\bf 42} (1990) 3567.
\bibitem{Allton:2008pn}
  C.~Allton {\it et al.}  [RBC-UKQCD Collaboration],
  Phys.\ Rev.\  D {\bf 78} (2008) 114509.

\bibitem{Alexandrou:1992ti}
  C.~Alexandrou, S.~Gusken, F.~Jegerlehner, K.~Schilling and R.~Sommer,
  Nucl.\ Phys.\  B {\bf 414} (1994) 815.

\bibitem{APEsmearing}
M. Albanese \emph{et. al.} (APE Collaboration) 
Phys. \ Lett. \ B {\bf 192} (1987) 163.


\bibitem{Alexandrou:2009hs}
  C.~Alexandrou {\it et al.},
  Nucl.\ Phys.\  A {\bf 825} (2009) 115.

\bibitem{Alexandrou:2007dt}
  C.~Alexandrou, G.~Koutsou, H.~Neff, J.~W.~Negele, W.~Schroers and A.~Tsapalis,
  Phys.\ Rev.\  D {\bf 77} (2008) 085012.

\bibitem{Aoki:2007xm}
  Y.~Aoki {\it et al.},
  Phys.\ Rev.\  D {\bf 78} (2008) 054510.

\bibitem{Boinepalli:2009sq}
  S.~Boinepalli, D.~B.~Leinweber, P.~J.~Moran, A.~G.~Williams, J.~M.~Zanotti and J.~B.~Zhang,
 Phys. Rev. D 80, 054505 (2009), arXiv:0902.4046 [hep-lat].

\bibitem{Aubin:2008qp}
  C.~Aubin, K.~Orginos, V.~Pascalutsa and M.~Vanderhaeghen,
 Phys. Rev. D 79, 051502 (2009), arXiv:0811.2440 [hep-lat].

\bibitem{Leinweber:1992hy}
  D.~B.~Leinweber, T.~Draper and R.~M.~Woloshyn,
  Phys.\ Rev.\  D {\bf 46} (1992) 3067.



\end{thebibliography}
\end{document}